\documentclass{elsart}
\usepackage[T1]{fontenc}
\usepackage[latin1]{inputenc}

\makeatletter

\providecommand{\LyX}{L\kern-.1667em\lower.25em\hbox{Y}\kern-.125emX\@}

\def\D0{D\O~}
\def\thisday{\today}

\def\beq{\begin{equation}}
\def\eeq{\end{equation}}
\def\bdm{\begin{displaymath}}
\def\edm{\end{displaymath}}
\def\bea{\begin{eqnarray}}
\def\eea{\end{eqnarray}}

\def\D0{D\0~}

\newcommand{\prescr}[2]{{\,}^{#1}{#2}}

\def\cal{}

\def\zh{\widehat z}

\def\Vh{\widehat{V}}

\def\url#1{\mbox{\href{#1}{\sf #1}}}

\usepackage{amssymb,graphicx}

\makeatother

\begin{document}

\begin{frontmatter}

\title{
Azimuthal asymmetries at HERA: theoretical aspects
}

\author{
P.M.~Nadolsky, D.R.~Stump,  C.--P. Yuan} \address{Department of Physics \& Astronomy,
Michigan State University,\\ East Lansing, MI 48824, U.S.A.}
\date{\thisday}
\begin{abstract}
We comment on theoretical aspects of the measurement of 
azimuthal asymmetries in semi-inclusive charged particle production, 
made recently by the ZEUS Collaboration at HERA.
By taking the ratio of the two measured asymmetries, we find good
agreement between the perturbative QCD prediction and the experimental
data. To separate the perturbative and nonperturbative 
contributions to the asymmetries, we suggest that the azimuthal 
asymmetries of the transverse energy flow be measured 
as a function of a variable \(q_{T}\) 
related to the pseudorapidity of the energy flow.
\end{abstract}

\begin{keyword}
Semi-inclusive deep inelastic scattering; QCD; resummation

\PACS
12.38.Bx \sep
12.38.Cy \sep
13.85.-t \end{keyword}
\end{frontmatter}

\vfill

\leftline{Preprint MSUHEP-01214; CTEQ-017; hep-ph/0012262}

\clearpage

\setcounter{footnote}{0} \renewcommand{\thefootnote}{\arabic{footnote}}

In a recent publication \cite{ZEUSasym} the ZEUS Collaboration at DESY-HERA
has presented data on asymmetries of charged particle (\( h^{\pm } \)) production
in the process \( e+p\stackrel{\gamma ^{*}}{\longrightarrow }e+h^{\pm }+X \),
with respect to the angle \( \varphi  \) defined as the angle between the lepton
scattering plane and the hadron production plane (of \( h^{\pm } \) and the
exchanged virtual photon). The azimuthal asymmetries, \( \langle \cos \varphi \rangle  \)
and \( \langle \cos {2\varphi }\rangle  \), as functions of the minimal transverse
momentum \( p_{c} \) of the observed charged hadron \( h^{\pm } \) in the
hadron-photon center-of-mass (hCM) frame, are defined as \begin{equation}
\label{chgdasym}
\langle \cos n\varphi \rangle (p_{c})=\frac{\int d\Phi \int _{0}^{2\pi }d\varphi \cos n\varphi \frac{d\sigma }{dxdzdQ^{2}dp_{T}d\varphi }}{\int d\Phi \int _{0}^{2\pi }d\varphi \frac{d\sigma }{dxdzdQ^{2}dp_{T}d\varphi }},
\end{equation}
 with \( n=1,2 \). In terms of the momenta of the initial proton \( P^{\mu } \),
the final-state hadron \( P_{h}^{\mu } \), and the exchanged photon \( q^{\mu } \),
the variables in (\ref{chgdasym}) are \( Q^{2}=-q_{\mu }q^{\mu } \), \( x=Q^{2}/2(P\cdot q) \),
and \( z=(P\cdot P_{h})/(P\cdot q) \). \( \int d\Phi  \) denotes the integral
over \( x,z,Q^{2},p_{T} \) within the region defined by \( 0.01<x<0.1 \),
\( 180\mbox {\, GeV}^{2}<Q^{2}<7220\mbox {\, GeV}^{2} \), \( 0.2<z<1,\, \mbox {and\, }p_{T}>p_{c} \).
Nonzero \( \langle {\cos {2\varphi }}\rangle  \) comes from interference of
the helicity \( +1 \) and \( -1 \) amplitudes of the transverse photon polarization;
and nonzero \( \langle {\cos \varphi }\rangle  \) comes from interference of
transverse and longitudinal photon polarization.

More than 20 years ago it was proposed to test QCD by comparing measured azimuthal
asymmetries to the perturbative predictions \cite{GeorgiPolitzer}. However,
it was also realized that nonperturbative contributions and higher-twist effects
may affect the comparison \cite{KKJK,ChgdNP,LeveltMulders94,BoerMulders98}.
For example, intrinsic \( k_{T} \) might be used to parametrize the nonperturbative
effects \cite{KKJK}, and indeed ZEUS did apply this idea to their analysis
of the data \cite{ZEUSasym}. The relative importance of the nonperturbative
effects is expected to decrease as \( p_{T} \) increases. Thus, the azimuthal
asymmetries in semi-inclusive deep-inelastic scattering (SIDIS) events with
large \( p_{T} \) should be accurately described by perturbative dynamics.
From the comparison to the perturbative QCD calculation at the leading order
in \( \alpha _{s} \) \cite{Gehrmann,KoppMendez}, the ZEUS Collaboration concluded
that the data on the azimuthal asymmetries at large values of \( p_{c} \),
although not well described by the QCD predictions, do provide clear evidence
for a perturbative QCD contribution to the azimuthal asymmetries.

In this paper, we will take a new look at the ZEUS data, motivated by a QCD
resummation formalism \cite{CSS,Meng2,nsy1999,nsy2000} that takes into account
the effects of multiple soft parton emission. First, we argue that the analysis
of \( \langle \cos \varphi \rangle  \) and \( \langle \cos {2\varphi }\rangle  \)
based on fixed-order QCD is unsatisfactory because it ignores large logarithmic
corrections due to soft parton emission. We also show that perturbative and
nonperturbative contributions are mixed in the transverse momentum distributions.
Then we make two suggestions for improvement of the analysis of the ZEUS data.
We show that perturbative and nonperturbative contributions can be separated
more clearly in asymmetries depending on a variable \( q_{T} \) related to
the pseudorapidity of the final hadron in the hCM frame. We also suggest measurement
of the asymmetries of the \emph{transverse energy flow} which are simpler and
may be calculated reliably. Our predictions for the asymmetries of transverse
energy flow are the most important contribution of this paper.

\section{Large logarithmic corrections and resummation}

The impact parameter resummation formalism that we are applying here describes
production of nearly massless hadrons in the current fragmentation region, where
the production rate is the highest. In this region, transverse momentum distributions
are affected by large logarithmic QCD corrections due to radiation of soft and
collinear partons. The leading logarithmic contributions can be summed through
all orders of perturbative QCD \cite{Meng2,nsy1999,nsy2000} by applying a method
originally proposed in \cite{CSS} for jet production in \( e^{+}e^{-} \) annihilation
and the Drell-Yan process. 

The spin-averaged cross section for SIDIS in a parity-conserving channel, \textit{e.g.,}
\( \gamma ^{*} \) exchange, can be decomposed into a sum of independent contributions
from four basis functions \( A_{\rho }(\psi ,\varphi ) \) of the leptonic angular
parameters \( \psi ,\varphi  \) \cite{Meng1}: \[
\frac{d\sigma }{dxdzdQ^{2}dq_{T}^{2}d\varphi }=\sum _{\rho =1}^{4}\prescr {\rho }{V}(x,z,Q^{2},q_{T}^{2})A_{\rho }(\psi ,\varphi ).\]
 Here \( \psi  \) is the angle of a hyperbolic rotation (a boost) in Minkowski
space; it is related to the conventional DIS variable \( y \), by \( y=P\cdot {q}/P\cdot {\ell }=2/(1+\cosh \psi ) \).
The angular basis functions are \( A_{1}=1+\cosh ^{2}\psi  \), \( A_{2}=-2 \),
\( A_{3}=-\cos \varphi \sinh 2\psi  \), \( A_{4}=\cos 2\varphi \sinh ^{2}\psi  \).
Of the four structure functions \( \prescr {\rho }V \), only \( \prescr {1}V \)
and \( \prescr {2}V \) contribute to the denominator of (\ref{chgdasym}),
\textit{i.e.,} the \( \varphi  \)-integrated cross section. Of these two terms,
\( \prescr {1}V \) is more singular and it dominates the rate. To explore the
singular contributions in \( \prescr {1}V \), we introduce a scale \( q_{T} \)
related to the \textit{polar} angle (\( \theta  \)) of the direction of the
final hadron in the hCM frame. A convenient definition is \begin{equation}
q_{T}=Q\sqrt{1/x-1}\, \exp {(-\eta )},
\end{equation}
 where \( \eta  \) is the pseudorapidity of the charged hadron in the hCM frame
(defined with respect to the direction of the momentum \( q^{\mu } \) of \( \gamma ^{*} \)).
In the limit \( q_{T}\rightarrow 0 \), the structure function \( \prescr {1}V \)
is dominated by large logarithmic terms; it has the form \( q_{T}^{-2}\sum _{k=1}^{\infty }(\alpha _{s}/\pi )^{k}\sum _{m=0}^{2k-1}v^{(km)}\ln ^{m}(q_{T}^{2}/Q^{2}) \),
where \( v^{(km)} \) are some generalized functions. To obtain a stable theoretical
prediction, these large terms must be resummed through all orders of perturbative
QCD. The other structure functions \( ^{2,3,4}V \) are finite at this order;
we approximate them by fixed-order \( {\cal O}(\alpha _{S}) \) expressions. 

In Eq.\,(\ref{chgdasym}), the numerator of \( \langle \cos {\varphi }\rangle  \)
or \( \langle \cos {2\varphi }\rangle  \) depends only on the structure function
\( \prescr {3}V \) or \( \prescr {4}V \), respectively. The measurement of
\( \langle \cos {\varphi }\rangle  \) or \( \langle \cos {2\varphi }\rangle  \)
must be combined with good knowledge of the \( \varphi  \)-integrated cross
section, \textit{i.e.,} the denominator of (\ref{chgdasym}), to provide experimental
information on the structure function \( \prescr {3}V \) or \( \prescr {4}V \).
Thus it is crucial to check whether the theory can reproduce the \( \varphi  \)-integrated
cross section as a function of \( p_{T} \) before comparing the prediction
for (\ref{chgdasym}) to the data. But, on the contrary, as shown in \cite{nsy2000},
the \( {\mathcal{O}}(\alpha _{s}) \) fixed-order cross section is significantly
lower than the data from \cite{ZEUSchgd96} in the range of \( p_{T} \) relevant
to the ZEUS measurements. This difference signals the importance of higher-order
corrections and undermines the validity of the \( {\cal \mathcal{O}}(\alpha _{s}) \)
result as a reliable approximation for the numerator of (\ref{chgdasym}). 

On the other hand, the resummation calculation \cite{nsy2000} with a proper
choice of the nonperturbative function yields a much better agreement with the
experimental data for the \( \varphi  \)-integrated \( p_{T} \)-distribution
from \cite{ZEUSchgd96}. One might try to improve the theoretical description
of the ZEUS data using resummation for the denominator of (\ref{chgdasym}).
However, the resummation calculation for \( d\sigma /(dxdzdQ^{2}dp_{T}d\varphi ) \)
in the phase space region relevant to the ZEUS data is currently not possible,
largely because of the uncertainty in the parameterization of the nonperturbative
contributions in this region. The resummed structure function \( \, ^{1}V \)
includes a nonperturbative Sudakov factor, which contains the effects of the
intrinsic transverse momentum of the initial-state parton and the nonperturbative
fragmentation contributions to the transverse momentum of the final-state hadron.
Without first determining this nonperturbative factor, \textit{e.g.,} from other
measurements, it is not possible to make a trustworthy theoretical prediction
for the denominator of (\ref{chgdasym}) and, hence, these azimuthal asymmetries.

The azimuthal asymmetries measured by ZEUS may also be sensitive to uncertainties
in the fragmentation to \( h^{\pm } \) in the final state. Indeed, the cross
section in (\ref{chgdasym}) includes convolutions of hard scattering cross
sections with fragmentation functions (FFs), integrated over the range \( 0.2<z<1 \).
Although the knowledge of FFs is steadily improving \cite{FFs}, there is still
some uncertainty about their \( z \)-dependence and flavor structure for the
range of \( Q \) relevant to the ZEUS measurement. Therefore the most reliable
tests of the theory would use observables that are not sensitive to the final-state
fragmentation. The asymmetries \( \langle \cos n\varphi \rangle  \) would be
insensitive to FFs if the dependence on the partonic variable \( \widehat{z} \)
were similar in the hard parts of the numerator and denominator of (\ref{chgdasym}),
so that the dependence on the FFs would approximately cancel. (We denote the
parton-level quantities by {}``\( \widehat{\quad \! \! } \){}''.) It is shown
in Appendix~B of \cite{nsy1999} that the partonic structure function \( \prescr {1}\Vh  \),
which dominates the denominator of (\ref{chgdasym}), contains terms proportional
to \( 1/\zh ^{2} \) that increase rapidly as \( \zh  \) decreases. However,
the most singular terms in the partonic structure functions \( \prescr {3,4}\Vh  \)
are proportional to \( 1/\zh  \). Therefore, the dependence on the FFs does
not cancel in the azimuthal asymmetries.

\begin{figure}[t] \begin{center} \includegraphics[height=7.5cm]{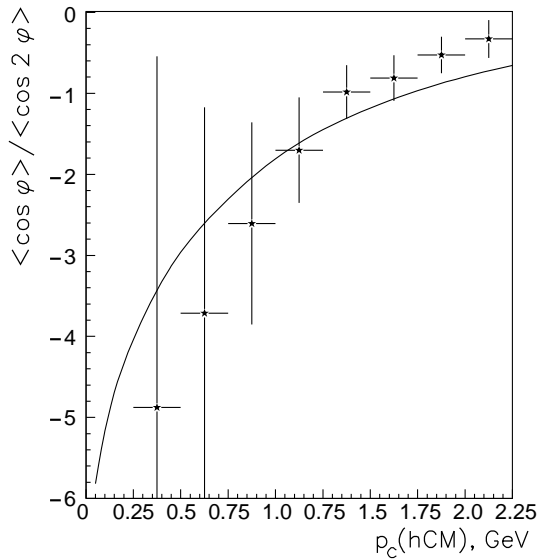} \caption{\label{cosphi2cos2phi}Comparison of the ${\cal O}(\alpha_s)$ prediction for the ratio $\langle \cos\varphi \rangle/\langle \cos{2\varphi} \rangle$ with the ratio of experimentally measured values of $\langle \cos\varphi \rangle$ and $\langle \cos{2\varphi} \rangle$ from \protect\cite{ZEUSasym}. The error bars are calculated by adding the statistical errors of $\langle \cos\varphi \rangle$ and $\langle \cos{2\varphi} \rangle$ in quadrature. Systematic errors are not included. The theoretical curve is calculated for $\langle x \rangle=0.022$, $\langle Q^2 \rangle=750\mbox{ GeV}^2$, using the CTEQ5M1 parton distribution functions \protect\cite{CTEQ5} and fragmentation functions by S.~Kretzer from \protect\cite{FFs}.\vspace{0.2cm}} \end{center} \end{figure}

A curious fact appears to support the suggestion that the theoretical predictions
for \( \langle \cos n\varphi \rangle  \) depend significantly on the fragmentation
functions. While each of the measured asymmetries, \( \langle \cos \varphi \rangle  \)
and \( \langle \cos 2\varphi \rangle  \), deviates from the \( {\mathcal{O}}(\alpha _{s}) \)
prediction, the data actually agree well with the \( {\mathcal{O}}(\alpha _{s}) \)
prediction for the ratio \( \langle \cos \varphi \rangle /\langle \cos 2\varphi \rangle  \),
as shown in Fig.\,\ref{cosphi2cos2phi}. The error bars are the statistical
errors on \( \langle {\cos \varphi }\rangle  \) and \( \langle \cos {2\varphi }\rangle  \)
combined in quadrature; this, however, may overestimate the statistical uncertainty
if the two errors are correlated. Since this ratio depends only on the numerators
in Eq.\,(\ref{chgdasym}), which are less singular with respect to \( \zh  \)
than the denominator, the dependence on the fragmentation functions may be nearly
canceled in the ratio. The good agreement between the \( {\mathcal{O}}(\alpha _{s}) \)
prediction and the experimental data for this ratio supports our conjecture
that the fragmentation dynamics has a significant impact on the individual asymmetries
defined in (\ref{chgdasym}). 


Our final remark about the azimuthal asymmetries in (\ref{chgdasym}) is that
the \( p_{T} \) (or \( p_{c} \)) distributions are not the best observables
to separate the perturbative and nonperturbative effects. The region where multiple
parton radiation effects are important is specified by the condition \( q_{T}^{2}/Q^{2}\ll 1 \).
But the \( p_{T} \) distributions are smeared with respect to the \( q_{T} \)
distributions by an additional factor of \( z \), because \( p_{T}=z\, \, q_{T} \).
Thus the whole observable range of \( p_{T} \) is sensitive to the resummation
effects in the region of \( q_{T} \) of the order of several GeV. A better
way to compare the data to the perturbative QCD prediction is to express the
azimuthal asymmetries as a function of \( q_{T} \), not \( p_{T} \). Then
the comparison should be made in the region where the multiple parton radiation
is unimportant, \textit{i.e.,} for \( q_{T}/Q\gtrsim 1 \).

\section{Asymmetry of energy flow}

Next, we describe an alternative test of perturbative QCD, which will further
reduce the above theoretical uncertainties: measurement of the azimuthal asymmetries
of the \emph{transverse energy flow}. In the hCM frame, the transverse energy
flow can be written as \cite{Peccei,Meng1,Meng2,nsy1999,nsy2000}\begin{equation}
\label{angET}
\frac{dE_{T}}{dxdQ^{2}dq_{T}^{2}d\varphi }=\sum _{\rho =1}^{4}\prescr {\rho }V_{E_{T}}(x,Q^{2},q_{T}^{2})A_{\rho }(\psi ,\varphi ).
\end{equation}

Unlike the charged particle multiplicity, the energy flow does not depend on
the final-state fragmentation. It has been demonstrated \cite{nsy1999,nsy2000}
that a resummation calculation can provide a good description for the experimental
data on the \( \varphi  \)-integrated \( E_{T} \)-flow. A new class of azimuthal
asymmetries may be defined as \begin{eqnarray}
\langle E_{T}\cos n\varphi \rangle (q_{T})=\frac{\int d\Phi \int _{0}^{2\pi }\cos n\varphi \frac{dE_{T}}{dxdQ^{2}dq_{T}^{2}d\varphi }d\varphi }{\int d\Phi \int _{0}^{2\pi }\frac{dE_{T}}{dxdQ^{2}dq_{T}^{2}d\varphi }d\varphi }.\label{ETasym} 
\end{eqnarray}
 The structure functions \( \prescr {\rho }V_{E_{T}} \) for the \( E_{T} \)-flow
can be derived from the structure functions \( \prescr {\rho }V \) for the
SIDIS cross section \cite{nsy2000}. Similar to the case of the particle multiplicities,
the asymmetries \( \langle E_{T}\cos \varphi \rangle  \) and \( \langle E_{T}\cos 2\varphi \rangle  \)
receive contributions from \( \prescr {3}V_{E_{T}} \) and \( \prescr {4}V_{E_{T}} \),
respectively. But, unlike the previous case, the denominator in (\ref{ETasym})
is approximated well by the resummed \( E_{T} \)-flow. Thus these asymmetries
can be calculated with greater confidence.

\begin{figure}[tbp] 

\hspace{-2cm}\includegraphics[width=17cm]{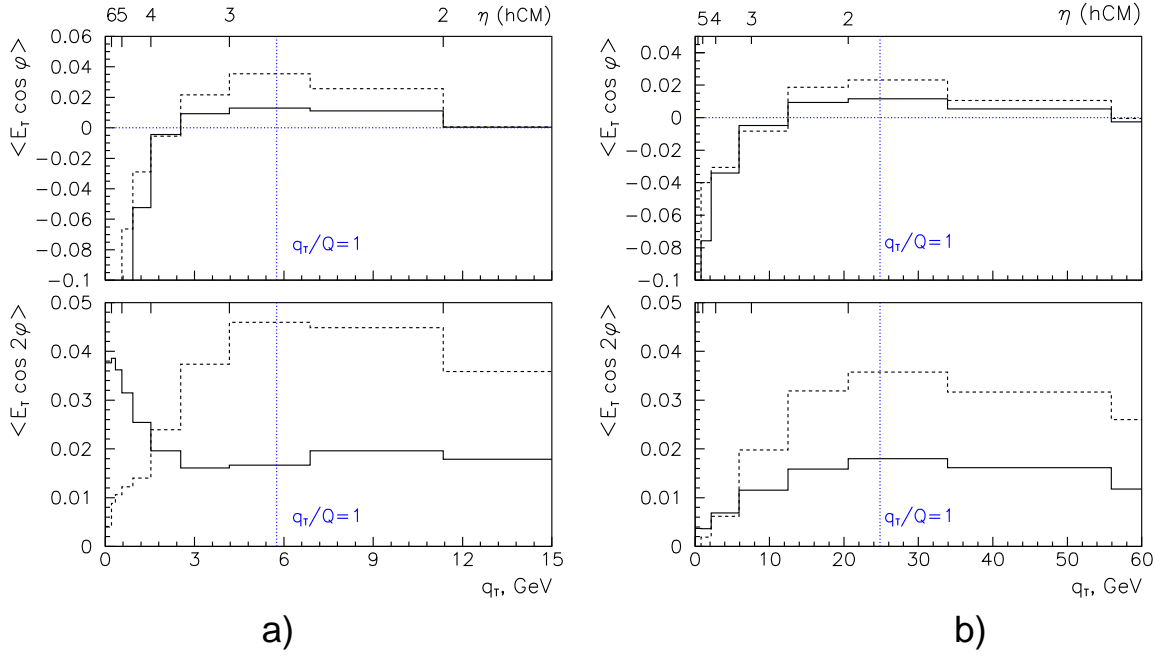} \caption{\label{ETasymqt} Energy flow asymmetries $\langle E_T \cos \varphi\rangle (q_T)$ and $\langle E_T \cos{2 \varphi} \rangle (q_T)$ for (a) $x = 0.0047$, $Q^2 = 33.2 \mbox{ GeV}^2$ and (b) $x = 0.026$, $Q^2 = 617 \mbox{ GeV}^2$. The Figure shows predictions from the resummed (solid) and the ${\cal O}(\alpha_s)$ (dashed) calculations.\vspace{0.2cm}} \end{figure}

Figure \ref{ETasymqt} shows our prediction for the azimuthal asymmetries \( \langle E_{T}\cos \varphi \rangle  \)
and \( \langle E_{T}\cos 2\varphi \rangle  \) as functions of \( q_{T} \)
for (a) \( x=0.0047 \), \( Q^{2}=33.2\mbox {GeV}^{2} \) in the left plots
and (b) \( x=0.026 \), \( Q^{2}=617\mbox {GeV}^{2} \) in the right plots.
The asymmetries are shown in \( q_{T} \)-bins that are obtained from the experimental
pseudorapidity bins for the \( \varphi  \)-integrated \( E_{T} \)-flow data
from Ref.~\cite{H1z2}. The upper \( x \)-axis shows values of the hCM pseudorapidity
\( \eta  \) that correspond to the values of \( q_{T} \) on the lower \( x \)-axis.
For each of the distributions in Fig.\,\ref{ETasymqt}, the structure functions
\( \prescr {3}V_{E_{T}} \) and \( \prescr {4}V_{E_{T}} \) were calculated
at leading order in QCD, \textit{i.e.,} \( {\mathcal{O}}(\alpha _{s}) \). The
solid and dashed curves, which correspond to the resummed and \( {\mathcal{O}}(\alpha _{s}) \)
results respectively, differ because the structure function \( \prescr {1}V_{E_{T}} \)
in the denominator of (\ref{ETasym}) differs for the two calculations. The
resummed \( \varphi  \)-integrated \( E_{T} \)-flow is closer to the data
than the fixed-order result, so that the predictions made by perturbative QCD
for the subleading structure functions \( \prescr {3}V_{E_{T}} \) and \( \prescr {4}V_{E_{T}} \)
will be confirmed if the experimental azimuthal asymmetries agree with the resummed
distributions.

A recent study \cite{nsy2000} shows that in the region \( q_{T}\sim Q \) the
resummed \( \varphi  \)-integrated \( E_{T} \)-flow is larger than the \( {\mathcal{O}}(\alpha _{s}) \)
prediction. This explains why the asymmetries for \( q_{T}\sim Q \) are smaller
for the resummed denominator than for the \( {\mathcal{O}}(\alpha _{s}) \)
denominator. In the region \( q_{T}/Q\ll 1 \), the asymmetries are determined
by the asymptotic behavior of the fixed-order and resummed \emph{partonic} structure
functions \( \prescr {\rho }\Vh _{E_{T}} \). As \( q_{T}\rightarrow 0 \),
the \( {\mathcal{O}}(\alpha _{s}) \) structure functions \( (\prescr {1}\Vh _{E_{T}})_{{\mathcal{O}}(\alpha _{s})} \),
\( \prescr {3}\Vh _{E_{T}} \), and \( \prescr {4}\Vh _{E_{T}} \) behave as
\( 1/q_{T}^{2} \), \( 1/q_{T} \) and \( 1 \), respectively. Thus, asymptotically,
the ratios \( \prescr {3,4}\Vh _{E_{T}}/(\prescr {1}\Vh _{E_{T}})_{{\mathcal{O}}(\alpha _{s})} \)
go to zero, although the \( q_{T} \) distribution for the asymmetry \( \langle E_{T}\cos \varphi \rangle  \)
is quite large and negative for small, but non-vanishing \( q_{T} \) (\textit{cf.}
Fig.\,\ref{ETasymqt}). Resummation of \( \prescr {1}\Vh _{E_{T}} \) changes
the \( q_{T} \)-dependence of the denominator, which becomes nonsingular in
the limit \( q_{T}\rightarrow 0 \). Consequently, the asymmetry \( \langle E_{T}\cos \varphi \rangle  \)
with the resummed denominator asymptotically grows as \( 1/q_{T} \) (\textit{i.e.,}
in accordance with the asymptotic behavior of \( \prescr {3}\Vh _{E_{T}} \)).
Hence neither the fixed-order nor the resummed calculation for \( \langle E_{T}\cos \varphi \rangle  \)
is reliable in the low-\( q_{T} \) region, so that higher-order or additional
nonperturbative contributions must be important at \( q_{T}\rightarrow 0 \).
The asymptotic limit for the resummed \( \langle E_{T}\cos 2\varphi \rangle  \)
remains finite, with the magnitude shown in Fig.\,\ref{ETasymqt}. Since the
magnitude of \( \langle E_{T}\cos 2\varphi \rangle  \) is predicted not to
exceed a few percent, an experimental observation of a large asymmetry \( \langle E_{T}\cos 2\varphi \rangle  \)
at small \( q_{T} \) would signal the presence of some new hadronic dynamics,
\textit{e.g.,} contributions from \( T \)-odd structure functions discussed
in \cite{BoerMulders98}.

Figure \ref{ETasymqt} shows that the predicted asymmetry \( \langle E_{T}\cos \varphi \rangle (q_{T}) \)
at \( q_{T}/Q=1 \) is about 1--2\% for the resummed denominator, while it is
about 2--4\% for the \( {\mathcal{O}}(\alpha _{s}) \) denominator. The asymmetry
\( \langle E_{T}\cos 2\varphi \rangle (q_{T}) \) at \( q_{T}/Q=1 \) is about
1.5-2\% or 3.5-5\%, respectively. Both asymmetries are positive for \( q_{T}\sim Q \).
According to Fig.\,\ref{ETasymqt}a, the size of the experimental \( q_{T} \)
bins (converted from the \( \eta  \) bins in \cite{H1z2}) for low or intermediate
values of \( Q^{2} \) is small enough to reveal the low-\( q_{T} \) behavior
of \( \prescr {3,4}V_{E_{T}} \) with acceptable accuracy. However, for the
high-\( Q^{2} \) events in Fig.\,\ref{ETasymqt}b, the experimental resolution
in \( q_{T} \) may be insufficient for detailed studies in the low-\( q_{T} \)
region. Nonetheless, it will still be interesting to compare the experimental
data to the predictions of perturbative QCD in the region \( q_{T}/Q\approx 1 \),
and to learn about the angular asymmetries at large values of \( Q^{2} \) and
\( x \).

To conclude, we suggest that the azimuthal asymmetry of the energy flow should
be measured as a function of the scale \( q_{T} \). These measurements would
test the predictions of the perturbative QCD theory more reliably than the measurements
of the asymmetries of the charged particle multiplicity.

\section*{Acknowledgments}

This work was supported in part by the NSF under grants PHY-9802564.

\end{document}